\documentclass[sigconf,letterpaper]{acmart}

\usepackage{graphicx} % Required for inserting images
\usepackage[english]{babel}
\usepackage{siunitx}
\usepackage{subcaption}
\usepackage{listings}
\usepackage{xcolor}
\definecolor{terminalbg}{rgb}{0.95, 0.95, 0.95} % Very light gray
\definecolor{terminalframe}{rgb}{0.7, 0.7, 0.7} % Medium gray for the border

\copyrightyear{2026}
\acmYear{2026}
\setcopyright{cc}
\setcctype{by}
\acmConference[IMC '26]{Proceedings of the 2026 ACM Internet Measurement Conference}{October 12--16, 2026}{Karlsruhe, Germany}
\acmBooktitle{Proceedings of the 2026 ACM Internet Measurement Conference (IMC '26), October 12--16, 2026, Karlsruhe, Germany}
\acmDOI{10.1145/3777912.3809153}
\acmISBN{979-8-4007-2327-8/2026/10}

\title{Characterizing the Configuration of Starlink Queuing }

\author{Johan Garcia}
\orcid{0000-0003-3461-7079}
\affiliation{
   \institution{Karlstad University}
   \city{Karlstad} 
   \country{Sweden} 
}
\email{johan.garcia@kau.se}

\author{Simon Sundberg}
 \orcid{0000-0002-3570-9525}
\affiliation{
   \institution{Karlstad University}
   \city{Karlstad} 
   \country{Sweden} 
}
\email{simon.sundberg@kau.se}

\author{Anna Brunstrom}
\orcid{0000-0001-7311-9334}
\affiliation{
   \institution{Karlstad University}
   \city{Karlstad} 
   \country{Sweden} 
}
\email{anna.brunstrom@kau.se}

\begin{abstract}
In all networking systems, queuing is important to ensure appropriate resource utilization in the presence of bursty traffic and varying traffic demands. 
The Starlink access network is additionally also dynamic in terms of the capacity it can provide, and thus queuing plays an even greater role to ensure appropriate communication performance for the end-users while maintaining high resource utilization. 
However, for Starlink most system design details, along with the setup of the internal queuing, is private information and not publicly available. 
To address this we have developed a high-precision, burst-pattern controlled, traffic generation approach allowing us to precisely measure the one-way delay 
for Starlink.
By analyzing the delay and loss in conjunction with a queue simulator we find 
that Starlink does not employ per-flow fair
queuing or drop-tail buffers, but it does use drop-front buffer
management. While drop-front reduces delay, 
it may also interfere with the assumptions made
by loss-based congestion controls, potentially contributing to
throughput degradation. 
\end{abstract}



\begin{document}

\maketitle

\section{Introduction}

Starlink currently provides Internet access to more than 8 million users~\cite{StarlinkStories}
and is an important connectivity provider to remote and under-served areas. Despite its importance in the Internet ecosystem as a connectivity provider with extremely wide coverage, many technical details of the Starlink network configuration, and the resulting performance, are not publicly known. While there have been a number of studies aiming to better understand the characteristics of various components of the Starlink network~\cite{pan2023measuring,pan2024measuring,garcia2023physRateInference,lanfer2024weather} and the service it provides~\cite{izhikevich2024global,mohan2024multifaceted,kassem2022browser,garcia2023multi}, so far there has been no study focusing on the queuing behavior of Starlink. The queuing behavior is of considerable importance, as it can have a large influence  both on latency, and on the congestion control (CC) behavior of the forwarded traffic. 

In this study, we perform a comprehensive examination of Starlink queuing utilizing two geographically separated user terminals. To have detailed  control of the generated probing traffic, a custom traffic generator is utilized. We design a probe pattern encompassing some 325000 UDP probe packets, which are grouped into 120 distinct traffic burst types, where the burst types vary in terms of burst send rate and burst length.
Through repeated experiments this results in more than 10 million probe packets, which are analyzed to infer details about the queuing configuration. We find evidence indicating that Starlink queues are not set up with per-flow fair queuing, but instead all flows of a user share a single queue. 
Perhaps most interestingly, we find that the buffer management strategy, in terms of deciding which packets to drop, is not the commonly used drop-tail, but rather a drop-front strategy. While drop-front reduces latency in that its average queuing delay is lower, drop-front may interact with CC dynamics.

This finding is quite significant, as it provides additional background to  the poor Starlink throughput performance observed by CCs that use packet loss as their main congestion signal. 
Severe underutilization of the potentially available communication resources occurs when a single flow is transferred using a loss-sensitive CC such as Cubic~\cite{kassem2022browser,lai2024mind,ma2023network,garcia2025tcp}.
As Cubic is the default CC in the Linux kernel, a large fraction of all traffic traversing Starlink can be expected to be Cubic traffic, and this traffic can attain considerably lower throughput possibly linked to  the queuing characteristics uncovered in this investigation.

This work provides the following main contributions:
\begin{itemize}
    \item A flexible queue characterization methodology employing precise pattern-controlled traffic generation.
    \item A queuing simulator allowing insights to be drawn by fitting against the empirical observations.
    \item A detailed characterization of the Starlink queuing configuration, including the likely queue size.
    \item Additional insight into the low-throughput anomaly that has been observed for loss-based CCs, such as Cubic, over the Starlink network.
\end{itemize}
To allow further analyses we provide our data and the queuing simulator developed for this research~\cite{zenodo_18868451}.

\section{Background and related work}
\label{sec:RelWork}

It is well known that the Starlink network characteristics vary over time in terms of available capacity, round-trip time and packet loss~\cite{garcia2023multi,lai2024mind, mohan2024multifaceted}. Previous work has established that Starlink uses a 15-second reconfiguration interval~\cite{garcia2023multi,mohan2024multifaceted} and schedules transmissions in 1.33 ms link layer frames~\cite{humphreys2023signal,garcia2023physRateInference}. The short term throughput of Starlink is dependent on the physical transmission rate, and the allocation structure of the link layer frames~\cite{garcia2024finegrained,garcia2025modeling}. 

Potential queuing within the user-facing edge router is examined in~\cite{hammer2024starlink}, but as this router is not the end-to-end bottleneck no significant buffer buildup was identified. In contrast, our work concerns the queue at the actual end-to-end bottleneck, and provides a detailed characterization of its behavior.
In terms of approaches for evaluating queuing configuration, the Tada approach~\cite{kargar2016tada} intends to detect if a bottleneck router uses active queue management (AQM) or not, while being robust to shared background traffic and potentially multiple bottlenecks. Our setting here is more constrained, while also having a wider analysis objective. A method for detecting FIFO, PIE and FQ-CoDel AQM in the context of  DASH-like streaming is described in~\cite{kua2020detecting}. Detecting the presence of fair queuing is covered in~\cite{bachl2020detecting}.

The queuing configuration interacts with the CC, and there have been several studies evaluating CCs specifically for Starlink. 
In~\cite{deutschmann2023internet} 10-flow measurements show that BBR performs better than Cubic,
while~\cite{lai2024mind} examines several CCs and finds that Cubic achieves a quarter of the uplink throughput compared to BBRv1. 
Other measurement studies include~\cite{mohan2024multifaceted}, which provides global throughput results based on 10s long single flow measurements from a large number of servers with unspecified CC configuration.
In~\cite{Hu23LEOvsCellular} TCP throughput is evaluated for single, four, and eight flows with an unspecified CC variant, likely Cubic. 
An early work~\cite{kassem2022browser} compared 5 CCs, including BBR and Cubic, using an unspecified number of flows, finding that BBRv1 achieves much higher throughput than Cubic. 
In~\cite{ma2023network} Iperf3 is used with Cubic, with an unspecified duration and number of flows. That work also measures throughput with UDP, and reports TCP achieving \qty{39}{\percent} of the UDP throughput, which indicates considerable underutilization for the Cubic CC. A recent comparison between Cubic, BBR and several other CCs highlighted that single flow Cubic on average reached less than a quarter of the BBR throughput~\cite{garcia2025tcp}.
There are also several studies focusing on LEO-specific CC improvements, including LeoCC~\cite{lai2025leocc} which also considers the impact of buffer management on their proposed scheme. LEO-adapted CC for QUIC is examined in~\cite{yang2024mobility,kamel2024starquic}, while SaTCP is proposed and evaluated against Cubic in~\cite{cao2023satcp}.

Some studies have used simulation or emulation to evaluate CC performance over Starlink ~\cite{jiang2023leotp,hervella2022realistic,khan2023realistic,barbosa2023comparative,lai2024mind, ohs2025phantomlink}. The correct queuing configuration is important for simulation and emulation fidelity, and can have considerable impact on CC performance. Thus, the results presented here will be beneficial when setting up simulation or emulation studies aiming to faithfully represent Starlink connectivity. 
Other measurement studies explicitly consider effects such as weather~\cite{lanfer2024weather, laniewski2024wetlinks, ma2023network}, the impact of mobility~\cite{beckman2024starlink, laniewski2024starlink,Hu23LEOvsCellular,ghafoori2024stars,ullah2025starlink,laniewski2025measuring}, the use of multi-connectivity~\cite{lopez2022empirical,beckman2024starlink,Hu23LEOvsCellular, lopez2023connecting}, CC for video streaming~\cite{izhikevich2024global,zhang2024starstream,fang2024robust}, and backbone/PoP aspects~\cite{pan2023measuring,pan2024measuring}.

\section{Measurement setup}

To collect measurements, we develop a new measurement utility, nanoprobe~\cite{nanoprobe}, %\footnote{The source code for nanoprobe together with the used probe schedule are available at \url{https://github.com/simosund/nanoprobe}.}, 
capable of recording hardware timestamps from the NIC for both incoming and outgoing packets. Nanoprobe is based on nanoping~\cite{nanoping}, a simple UDP ping client and echo server with support for hardware timestamps. However, nanoprobe includes extensive modifications to improve the accuracy of the targeted inter-packet delay. We have previously used nanoprobe to measure and model the baseline one-way delay (OWD) of Starlink using low rate probing traffic~\cite{garcia2025detailed}. A novel feature is the nanoprobe support for sending traffic according to a probe schedule, a functionality not provided by previous tools such as IRTT~\cite{irtt}. In the probe schedule, each packet transmission is defined in terms of the number of microseconds of delay relative to the previous packet and the packet size for that probe packet. This functionality allows for very flexible generation of probing traffic. For our queuing experiments, we define a specific probe schedule where each measurement run consists of 120 traffic bursts, each with different 
burst characteristics in terms of 
number of packets and send rate within the burst. 

Our measurement campaign collects OWD data for traffic in the downlink in May 2025. 
Each 120-burst measurement run is replicated 30 times
to allow for the capture of aggregate behavior and account for Starlink's inherent variability.
Our primary set of measurements is performed from a Gen-2 Starlink Standard Actuated user terminal mounted on a rooftop at Karlstad University, Sweden, offering an unobstructed view of the sky. We utilize both software timestamps with tcpdump, and nanoprobe hardware timestamps from the dual-port Intel X550 NIC. The nanoprobe client is attached to a NIC port connected to the Starlink user terminal, while the server is attached to the second NIC port, which has a public IP address on the university network. We avoid clock drift and other issues with clock synchronization as both the client and the server rely on a common oscillator. The traffic is carried over a well-provisioned academic network and then a large international carrier (Twelve99) to the Starlink point of presence (PoP) in Frankfurt, where it enters the Starlink backbone to be forwarded on to a ground station and the LEO segment. The approximate non-Starlink delay between our measurement server and client is 12 ms.

\begin{figure*}[tp]
    \centering
    \begin{subfigure}{0.49\textwidth}
        \includegraphics[width=0.95\textwidth]{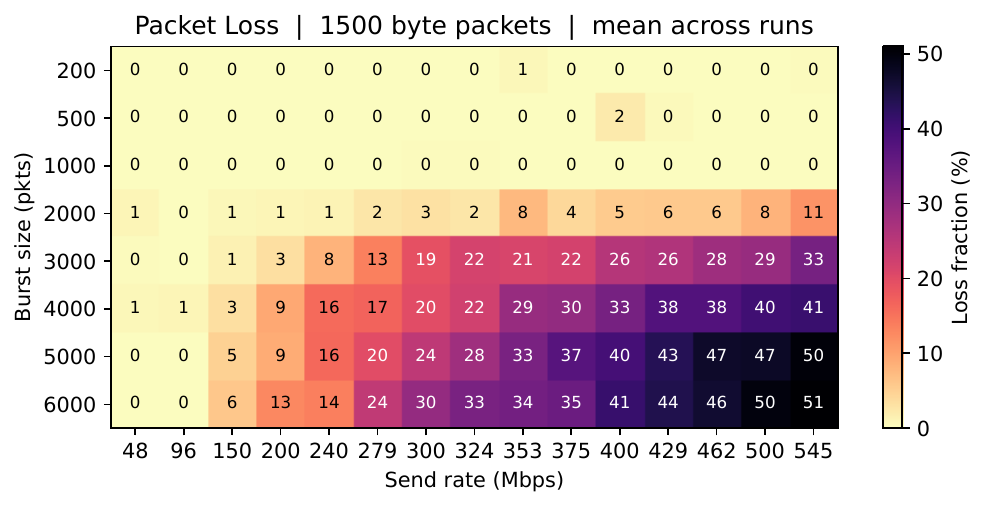} 
        \caption{Packet-loss}
        \label{fig:burst-PackLoss}
    \end{subfigure}
    \hfill
    \begin{subfigure}{0.49\textwidth}
        \includegraphics[width=0.95\textwidth]{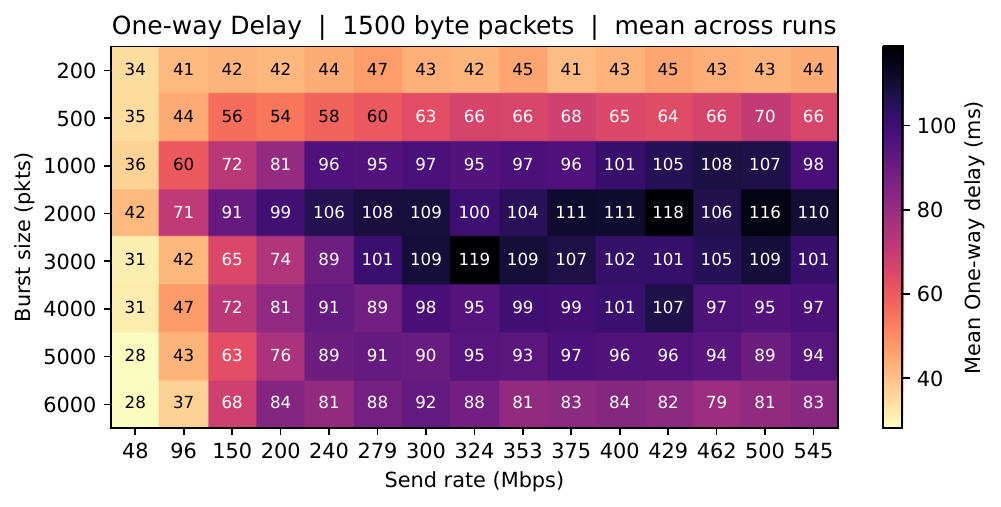}
        \caption{One-way delay}
        \label{fig:burst-OWD}
    \end{subfigure}

    \caption{Packet loss and one-way delay across 30 replications of burst size and burst send rate combinations.}
    \label{fig:pktloss_owd_heatmap}
\end{figure*}

Additionally, we repeat the measurements in October 2025 at a second vantage point with a Starlink terminal at the University of Malaga, Spain. 
The Malaga setup also runs both the client and the server on the same host, but 
relies solely on software timestamps.
The network path
is also similar, with traffic going through an academic network and then the Twelve99 network before reaching the Starlink PoP in Madrid.
Traceroute printouts for both the Karlstad and Malaga routes are provided in Appendix~\ref{sec:traceroute}.
The analysis is primarily based on the more precise measurements from Karlstad, while the Malaga measurements are used to verify that the observed behavior generalizes to other Starlink connections.

\section{Per-flow or per-user queues}

We initially study the queuing setup from a flow-separation viewpoint, and find that Starlink does not use per-flow fair queuing. 
To arrive at this result, some specific measurements are performed using TCP and Iperf3. First, 10 primary TCP flows with Cubic as CC are started. After 30 seconds,  10 additional TCP flows with the BBRv1 CC are started, and the 20 flows then run concurrently for the last 30 seconds. These measurement flows carry no features that would allow Starlink to directly distinguish between BBR and Cubic flows.
Varying radio conditions, varying intensity of traffic from other users, and scheduling considerations can all contribute to variations in the Starlink capacity available to a user, or here, to the measurement flows.
However, if per-flow queuing is present, each different flow will be allocated to its own queue, and a typical per-flow configuration is for each queue to be served an equitable fraction of the overall capacity over time. 
In such a case, one of the flows will thus not be able to, over time, capture a substantially larger fraction than any competing capacity-seeking flow. Consequently, per-flow fair queuing with round-robin allocation will enforce approximate max-min fairness among the flows. 

\begin{figure}[h!]
    \centering
    \includegraphics[width=0.99\columnwidth]{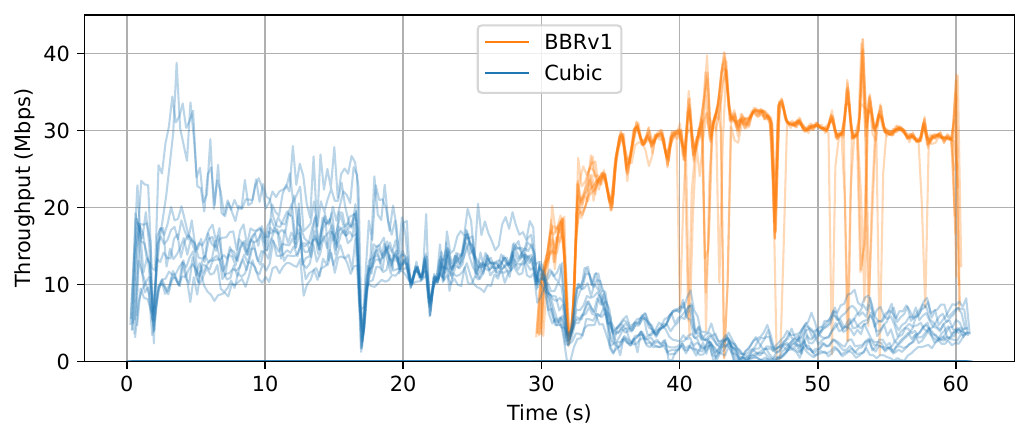}
    \caption{Per-flow throughput for 10 parallel Cubic and BBRv1 flows. BBRv1 flows start at 30 s.}
    \label{fig:tcp_fairness}
\end{figure}

If we now consider the throughput evolution graphs for these competing TCP connections as shown in Figure~\ref{fig:tcp_fairness}, they suggest that such per-flow fair queuing is not employed in Starlink.  In the later part of the transfer, it is clear that BBR takes a much larger fraction than Cubic, and that Cubic has a much lower throughput in the later part than in the first part of the transfer. Such behavior is not compatible with max-min fairness, and thus we can assume that per-flow fair queuing is not used.

\begin{figure*}
    \centering
    \begin{subfigure}{0.31\textwidth}
        \includegraphics[width=0.99\textwidth]{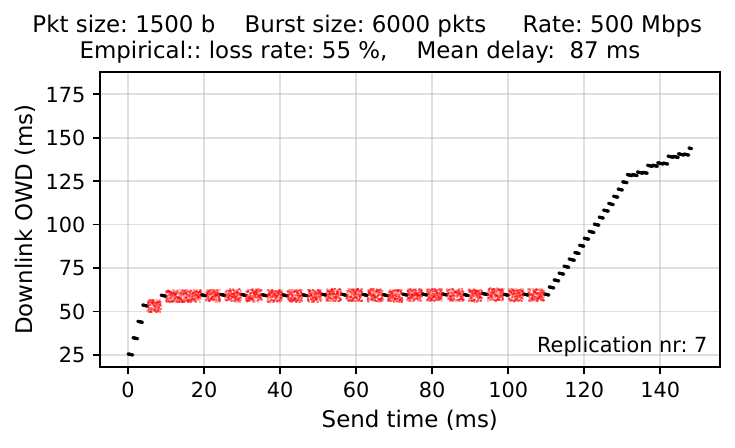}
        \caption{One-way delay and loss}
        \label{fig:sub-owdevo}
    \end{subfigure}
    \hfill
    \begin{subfigure}{0.31\textwidth}
        \includegraphics[width=0.99\textwidth]{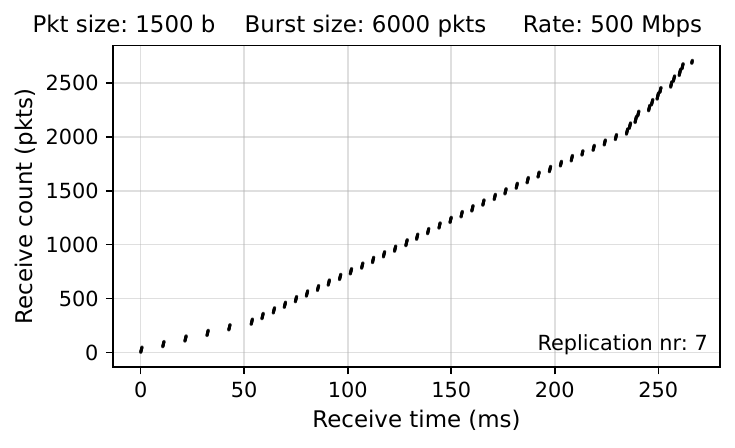}
        \caption{Received packet count}
        \label{fig:sub-recvPktCount}
    \end{subfigure}
    \hfill
    \begin{subfigure}{0.31\textwidth}
        \includegraphics[width=0.99\textwidth]{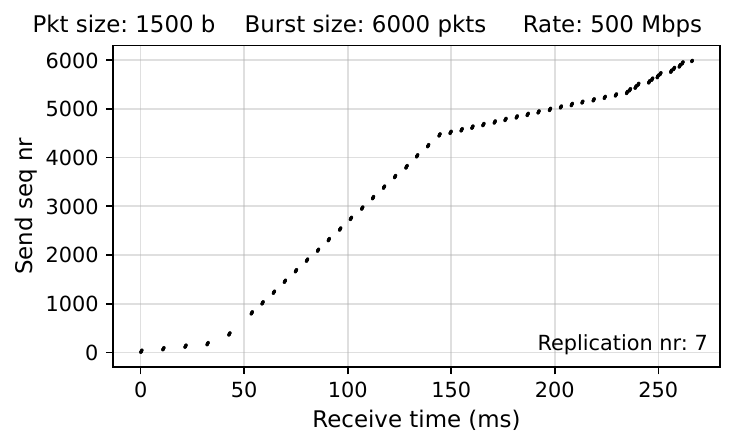} 
        \caption{Received sequence number}
        \label{fig:sub-recvSeqnr}
    \end{subfigure}
    \caption{Queue and packet receive dynamics for one illustrative Starlink burst measurements run with a burst size of 6000 (UDP) packets of 1500 bytes, and a constant send rate of 500 Mbps. Jittered red marks are packet losses. }
    \label{fig:example_OWD_timeseries}
\end{figure*}

\begin{figure*}
    \centering
    \begin{subfigure}{0.31\textwidth}
        \includegraphics[width=0.99\textwidth]{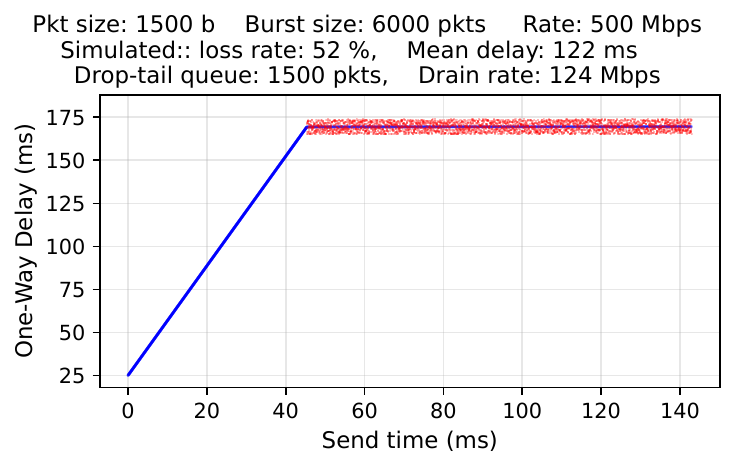} 
        \caption{Drop-tail}
        \label{fig:sim-droptail}
    \end{subfigure}
    \hfill
    \begin{subfigure}{0.31\textwidth}
        \includegraphics[width=0.99\textwidth]{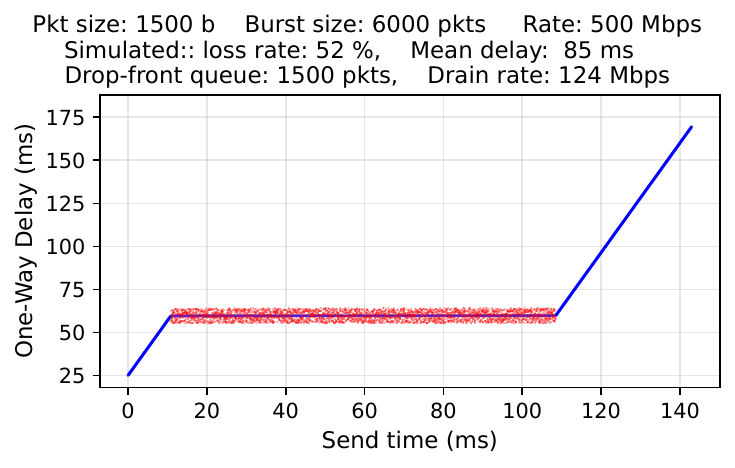}
        \caption{Drop-front}
        \label{fig:sim-dropfront}
    \end{subfigure}
    \hfill
    \begin{subfigure}{0.31\textwidth}
        \includegraphics[width=0.99\textwidth]{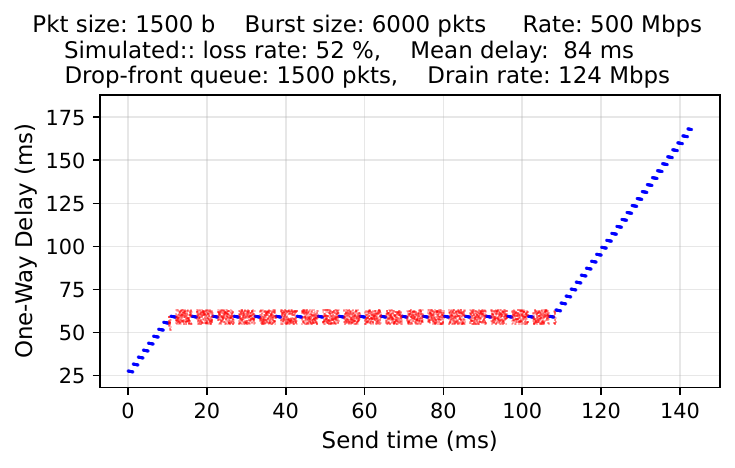} 
        \caption{Drop-front, Starlink drain}
        \label{fig:sim-dropfrontsl}
    \end{subfigure}
    
    \caption{One-way delay and loss dynamics obtained from queuing simulator for different configurations.}
    \label{fig:simulator_tail_vs_front_drop_OWD_timeseries}
\end{figure*}

\section{Queuing behavior overview}
\label{QbehaviorOverview}

To provide an overview of queuing dynamics we present heat maps  for the loss rate and the mean within-burst  one-way delay (OWD) in Figure~\ref{fig:pktloss_owd_heatmap}, based on our UDP burst-probe measurements. The heat maps show the mean behavior over the 30 replications for each of the 120 combinations of burst size and burst send rate. While the mean behavior as shown in the figure is representative, there are also individual variations as will be exemplified  later.

Several noteworthy aspects are present in these heat maps. Considering the loss fraction shown in Figure~\ref{fig:burst-PackLoss}, it can be observed that we have practically zero loss for burst sizes of 1,000 packets or less, regardless of the sending rate. This gives us an indication for a lower bound of the queue size. Furthermore, we can see that for burst rates of 48 and 96 Mbps, there is also practically no loss regardless of the burst size. This indicates that the available capacity on Starlink for our location at the time of the UDP measurement was lower-bounded by a value not too far from 96~Mbps. 
In the remaining part of the loss rates subplot, it is clear that the loss rate increases proportionally as the send rate and the burst size increase. This is consistent with the expected behavior of a queuing system where the incoming rate exceeds the drain  rate.

Additional insights can be gained by studying Figure~\ref{fig:burst-OWD} which shows the within-burst mean OWD. First, considering the low send rates of 48 and 96 Mbps, which had no packet losses, we can see that for 48 Mbps, there is a small tendency of increased latency when going from burst sizes of 200 to 2,000, and then a reduction in mean OWD as the burst sizes increase further. This trend can also be seen for the 96 Mbps send rate. Given that no losses were occurring for these send rates, one plausible explanation is that the underlying communications capacity that Starlink provides is sometimes increased for the longer bursts, thus driving down the mean OWD over the full burst. This indicates that Starlink can employ load-based allocation of the transmission resources, where a user initially gets some allocation of the 1.33 millisecond Starlink link layer frames that results in a particular rate. If the sender continuously transmits above this rate, Starlink can respond by increasing the resource allocation to the user, thus providing more capacity.

Considering the remainder of the OWD graph, it can be noted that the overall trend is different to what is seen for the loss rate, where losses increased corresponding to increases in send rate and burst size. For the OWD graph, the maximum delay instead appears for burst sizes of 2,000 packets and then actually reduces as the burst size increases. One potential partial explanation for this could be the load-dependent resource allocation discussed above, where larger bursts receive greater capacity in the later part of the burst, thus driving down the OWD.
However, a more significant explanation for this behavior is the particular drop-front buffer management scheme used, as discussed next.

\section{Buffer management}

In terms of buffer management approaches, we find that Starlink uses drop-front as opposed to the commonly employed drop-tail approach. 
To initiate the study of queue dynamics, and exemplify from the aggregated results of Figure~\ref{fig:pktloss_owd_heatmap}, we consider the behavior within one individual illustrative burst measurement. 
Figure~\ref{fig:sub-owdevo} shows the evolution of the per-packet OWD on the y-axis, as related to the send time on the x-axis. Within a burst, all packets are sent with a fixed inter-packet distance, which corresponds to a particular constant sending rate. From the figure it is clear that the OWD initially increases. As the queue becomes full, packet losses start to occur, as marked with red marks with added y-axis jitter for clarity. Between the losses, there are also packets that are forwarded to the client.
At the end of the transmission, there is an increase in the OWD, which is typical of a drop-front queue.
We can see that the send time duration is somewhat larger than 140 milliseconds. This time corresponds to  the particular burst size and burst sending rate that were used 
for this particular burst measurement run, i.e. $1500 \times 8 \times 6000 / 500\cdot
10^6 \lessapprox 0.144 $ seconds.

Figure~\ref{fig:sub-recvPktCount} provides an alternative view of the queue and link dynamics. It shows the client perspective and illustrates the effective reception rate, i.e. the effective capacity provided by Starlink. The y-axis shows the number of received packets, while the x-axis shows the reception time in relation to the first received packet.
From the y-axis numbering, it is clear that a bit more than 2500 packets were received out of the 6000 packets sent in this particular burst. For the given burst size and send rate, the loss fraction observed in this particular replication $(1 -(2711/6000) \approx 55 \% )$ can be compared to the mean across all replications (50 \%) from Figure~\ref{fig:pktloss_owd_heatmap}. 
This particular replication thus has a slightly higher loss rate compared to the mean over all replications.

\begin{figure*}
    \centering
    \begin{subfigure}{0.31\textwidth}
        \includegraphics[width=0.99\textwidth]{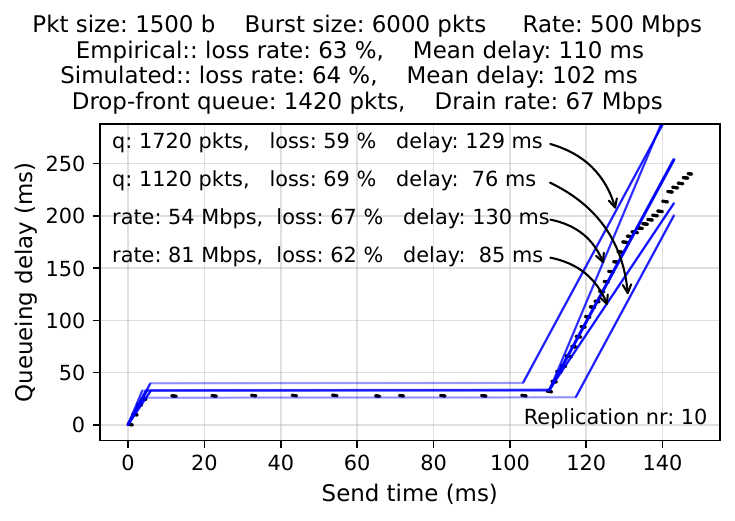}
        \caption{Simulator parameter sensitivity}
        \label{fig:sub-simparamsens}
    \end{subfigure}
    \hfill
    \begin{subfigure}{0.31\textwidth}
        \includegraphics[width=0.99\textwidth]{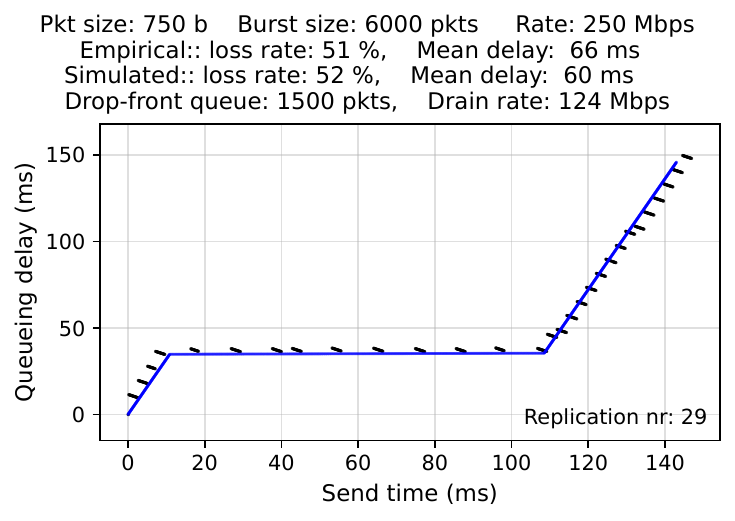}
        \caption{Smaller packet size}
        \label{fig:sub-smallPkt}
    \end{subfigure}
    \hfill
    \begin{subfigure}{0.31\textwidth}
        \includegraphics[width=0.99\textwidth]{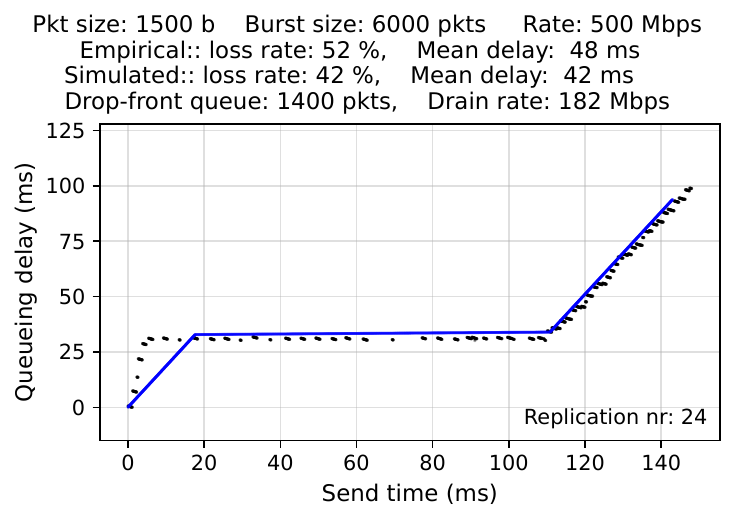} 
        \caption{From different location}
        \label{fig:sub-diffLocation}
    \end{subfigure}
    \caption{Per-packet measured queuing-induced OWD for 3 different runs (black), and per-packet OWD from a drop-front queuing simulator configured with queue size and drain rate to fit empirical measurements (blue). }
    \label{fig:examples_vs_simulator_OWD_timeseries}
\end{figure*}

Within the graph, the slope of an imagined interpolated line between the received small bursts of packets, corresponds to the received traffic rate, i.e., the provided Starlink capacity. In this particular run, we can see that initially the rate is lower, then it increases around the 50 ms receive time mark, and increases again at approximately 230 ms.
Thus, in this particular run, the provided underlying Starlink capacity was adjusted two times.

In Figure~\ref{fig:sub-recvSeqnr}, the data is again viewed from the client perspective. 
However, the y-axis now shows the sequence number of the packet that is received, rather than just the count of packets. The shape of the graph is similar to Figure~\ref{fig:sub-recvPktCount}, with some notable differences in terms of the slope. Here, the slope becomes steeper after the 50 millisecond mark. This is due to the packet losses that occur in this period, where the packets actually received will have a stepwise increase in their sent sequence number corresponding to the lost packets. 
There is also a noteworthy kink in the slope located slightly before the 150 millisecond mark. This kink appears at the same time as the sending of packets stops, which is slightly after 140 milliseconds, as seen in Figure~\ref{fig:sub-owdevo}. There is also a second kink around the 230 millisecond mark, which is caused by the capacity-adjustment change seen in Figure~\ref{fig:sub-recvPktCount}, although here it is not as visually pronounced. 

To further analyze 
the OWD evolution observed in Figure~\ref{fig:sub-owdevo}, we implemented a time-slot based Python queuing simulator with 1 $\mu$s resolution. The simulator takes traffic burst, queue, and drain characteristics as input, and generates traces of the resulting per-packet delay and loss for the configured drop strategy. We implemented the common drop-tail and the more rare drop-front strategies. 
Figure~\ref{fig:sim-droptail} shows the OWD evolution from the simulator for the same burst characteristics as shown in Figure~\ref{fig:sub-owdevo} when using drop-tail. This can be compared to the drop-front OWD evolution shown in Figure~\ref{fig:sim-dropfront}, which clearly provides a much better macroscopic match to Figure~\ref{fig:sub-owdevo}. Note that the configured queue size and drain rate are the same in \ref{fig:sim-droptail} and \ref{fig:sim-dropfront}, as is the resulting loss rate. The resulting mean delay is, however, considerably lower for drop-front. 
This is because with drop-front, packets are dropped from the front of the queue, meaning that the dropped packets are the ones which have waited the longest in the queue. This also results in the losses starting at a lower send time as compared to the drop-tail case. 
Notable is that when a packet is dropped at the front, the other packets in the queue are moved forward, thus lowering the total queuing delay for the packets that end up not being dropped at the front but delivered to the receiver. The difference in mean OWD provides a measure of the average effect on latency, and it is clear that drop-front is preferable from the perspective of keeping the queuing delay, and thus the OWD, low. As seen, the mode of the queuing-induced OWD is several times larger for drop-tail than for drop-front in this example run.
Interestingly,  for the drop-front case, the OWD will actually increase for the shorter bursts compared to longer bursts. While initially counterintuitive, this makes sense as the last phase where OWD has a rising slope (i.e. where the queuing delay is no longer reduced by the packet drops), is relatively longer. 
Thus, part of the explanation for the Figure~\ref{fig:pktloss_owd_heatmap}  behavior of decreasing OWD with larger bursts clearly lies with drop-front buffer management.  

In Figure~\ref{fig:sim-dropfrontsl} we further extended the simulator to not use smooth draining, but instead mimic the Starlink-specific 1.33 ms frame-time resource allocation, here with 1 frame-time with access, followed by 3 frame-times without access. This 1:3 pattern of allocation has been observed in previous empirical measurements~\cite{garcia2025modeling}.
The resulting matching of microscopic behavior in terms of loss placement between the empirical results in Figure~\ref{fig:sub-owdevo} and the simulated Starlink-specific results in Figure~\ref{fig:sim-dropfrontsl} shows 
that the observed buffering occurs at the queue in front of the Starlink resource-allocating bottleneck, and not somewhere else along the end-to-end path. 
If queuing, and thus loss generation, were to occur in a queue before Starlink, there would be losses also during the Starlink send-burst periods as the draining of a pre-Starlink queue would be independent of the Starlink send periods. So while we can conclude that queuing happens in a queue which is synchronized with Starlink's 1.33 ms frame time, it is not possible to pinpoint the exact location of this queue since there is very limited IP layer visibility into the internal Starlink network. Although there have been some recent efforts~\cite{wu2025beneath} to study the Starlink terrestrial network, many details remain unknown.

\begin{figure*}
    \centering
    \includegraphics[width=0.32\textwidth]{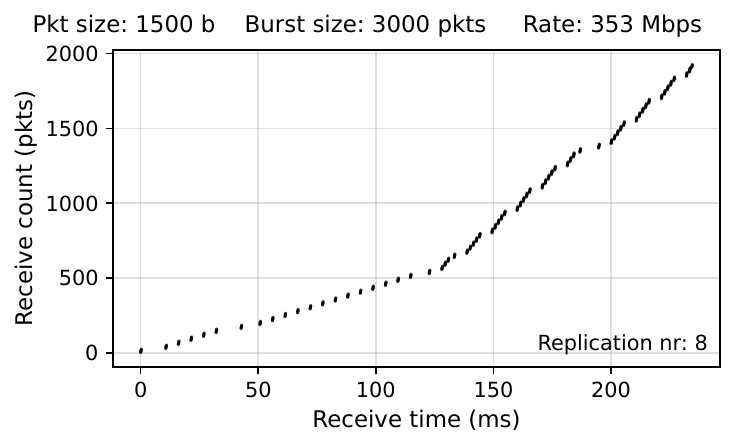}
    \hfill
    \includegraphics[width=0.32\textwidth]{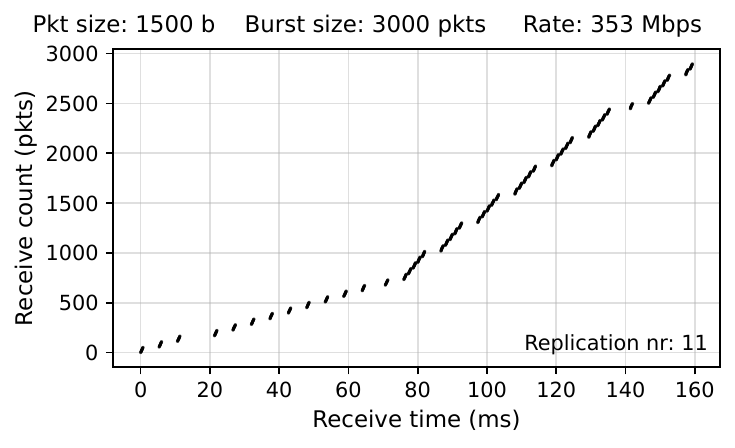}
    \hfill
    \includegraphics[width=0.32\textwidth]{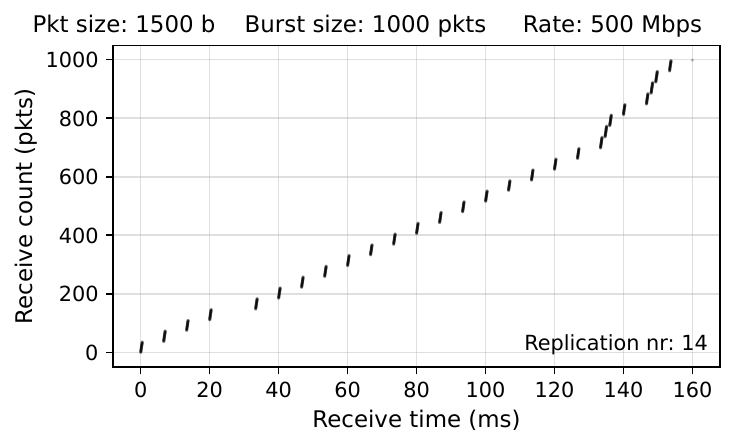}
    \caption{In-burst rate increase examples from two different burst sizes. Steeper slope indicates higher received rate.  }
    \label{fig:examples_burstrate_timeseries}
\end{figure*}

There are however also potential drawbacks to using drop-front buffer management. Cubic is the default congestion control (CC) in Linux, and as such it is widely deployed across Internet servers.  Many loss-based CCs such as Cubic are designed with the assumption of the queue being drop-tail, or some drop-tail based variation. A drop-front behavior will send the congestion signal earlier to the sender, possibly leading to underutilization of the available capacity. For Starlink, such underutilization has been reported in several works~\cite{deutschmann2023internet,lai2024mind,ma2023network}.
Another potential issue with drop-front is that it tends to synchronize multiple flows. If one flow feeds a high-rate burst of ten packets into a full drop-tail queue, the resulting 10 packet losses will  affect only itself. For the drop-front case, the ten packets will be fed into the queue and instead ten packets at the front of the queue will be dropped. It is highly likely that these packets will not all belong to the same flow that generated the high-rate burst. More likely, the ten packets will belong to some wider subset of all the competing flows, and this common loss occurrence will serve as a nudge towards flow synchronization for these flows. Thus, in many cases it will not be the flow that caused a packet loss that actually receives the congestion signal.

\section{Queue sizing}

By leveraging our queuing simulator we also investigate queue sizing, and find the typical queue size to be in the 1400 to 1600 packet range.
Three example burst measurements are shown in Figure~\ref{fig:examples_vs_simulator_OWD_timeseries}. 
The graphs are adjusted with the minimum observed OWD to display only the queuing-induced OWD increase, and are shown without packet losses. 
For figure clarity, we employ the smooth rather than the Starlink-specific draining configuration. 
In Figure~\ref{fig:sub-simparamsens} we provide an illustration of the impact of varying the  simulator parameters for queue size or drain rate when searching for a fit between empirical results and simulator output. For the fits shown in this paper, we performed manual fitting as no suitable and robust error metric could be immediately determined, and we thus leave automated fitting to future work. 
Figure~\ref{fig:sub-smallPkt} instead considers an auxiliary measurement run with traffic bursts using a smaller packet size. This run, and the corresponding fit, showed a similar queue size to the other runs with 1500 byte packets. This indicates that the queue is limited in terms of packets rather than bytes. If the queue had a byte limitation, then the fitted queue size value would be expected to be approximately double when the packet size is halved.
In Figure~\ref{fig:sub-diffLocation} we show a result obtained from measurements performed at Malaga, which is located  several thousand kilometers away from 
Karlstad. 
While these measurement locations varied in terms of obtained throughput and latency, they both exhibited the typical drop-front OWD evolution and similar queue size. Thus, these measurements confirm that the observed behavior is Starlink-specific, and not an artifact of some specific measurement setup or end-to-end path.

Additional examples of measurement runs and simulator fits are provided in Figure~\ref{fig:examples_simfits} of Appendix~\ref{sec:additionalRuns}. As the displayed runs were randomly selected, they can be seen to provide a representative view of the amount of between-run variation present in our measurements. While the simulator parameters leading to the best fit typically will not yield identical values for loss rate and mean OWD as empirically observed, the fit is nevertheless quite good. Also, note that the simulator does not model the rate adjustments that may occur in the empirical data.
From the perspective of queue sizing, the fitted parameters of the queuing simulator indicate a typical queue size of 1,400 to 1,600 packets. 
The variation in the OWD values at the plateaus in the different graphs goes from approximately 30 to 60 ms,  suggesting that drop decisions are related to queue size and not CoDel-like~\cite{rfc8289} delay bounds.

\section{Load-based rate adaptation}

As discussed in Section~\ref{QbehaviorOverview}, the decreasing mean OWD for the non-lossy burst rates of 48 and 96 Mbps indicated that there could be demand-based rate adaptation. 
In Figure~\ref{fig:examples_burstrate_timeseries}, we provide three additional receive count graphs using two smaller burst sizes: 3,000 and 1,000 packets. In these examples, it is clear that rate adaptation covers a larger fraction of all packets when the burst size is larger, whereas for the smaller burst size, rate adaptation covers a smaller fraction of all packets. 
With the rate adaptation affecting a larger proportion of the packets in large bursts, the average OWD will consequently be lower for large bursts, as previously observed and discussed for Figure~\ref{fig:pktloss_owd_heatmap}.

When examining all replications across a range of burst sizes and send rates, we do not find consistent evidence supporting load-based rate adaptation in a systematic and deterministic way. Nevertheless, rate adaptation appears in a number of runs and is evident both in the runs shown in Figures~\ref{fig:example_OWD_timeseries} and~\ref{fig:sub-simparamsens} and in many of the runs in Figure~\ref{fig:examples_simfits} of Appendix~\ref{sec:additionalRuns}, where some additional discussion on rate adaptation is also provided. Note that for the receive count graphs shown here, an increased drain rate from the queue is visible as a steeper slope, while in queuing delay graphs it is visible as a shallower slope.

\section{Conclusions}

Starlink is an important Internet access provider to millions of users globally, but many aspects of its underlying network configuration are poorly understood. In this study, we utilize a pattern-controlled traffic generator to characterize the queuing configuration of Starlink through the transmission and analysis of more than 10 million probe packets over a range of burst sizes and burst rates. By employing a queuing simulator we find that drop-tail buffering is not used, but rather a slightly unusual drop-front buffer management approach on a queue with a size of around 1500 packets. While drop-front reduces one-way delay and thus latency, it may interfere with the assumptions made by loss-based congestion controls. Our analysis also shows that Starlink does not employ per-flow fair queuing.
We believe that our  work contributes with relevant tools and methodology, and to additional insight into the causes  behind the Starlink throughput anomaly  that has been observed for Cubic and other loss-based congestion controls in several recent Starlink studies. However, this work cannot in itself establish a causal link between drop-front buffer management and poor Cubic performance, prompting future work on this throughput anomaly. We note that SpaceX may change the configuration of the network components, including the queue configuration, over time and has reported doing so~\cite{starlink2024latency}. 
We are currently working on a more complete link characterization measurement setup that can detect and quantify such changes.

\begin{acks}
Thanks to Jonas Karlsson at Karlstad and Delia Rico at Malaga for assisting with the measurement infrastructure.
This research was partly funded by the \grantsponsor{VR}{Swedish Research Council}{}, the \grantsponsor{KK}{Swedish Knowledge Foundation}{}, and \grantsponsor{KIPL}{Karlstad Internet Privacy
Lab (KIPL)}{}.
\end{acks}

\bibliographystyle{ACM-Reference-Format}
\bibliography{2406-DRIVE-StarlinkMultiCCeval}

@misc{nanoprobe,
  author = {Sundberg, Simon},
  title        = {nanoprobe: {UDP} packet generator with hardware timestamp support based on nanoping},
  year         = {2026},
  publisher    = {GitHub},
  journal      = {GitHub repository},
  howpublished = {\url{https://github.com/simosund/nanoprobe}},
}

@dataset{zenodo_18868451,
  author       = {Garcia, Johan and Sundberg, Simon},
  title        = {{Data set for "Characterizing the Configuration of Starlink Queuing",}},
  month        = mar,
  year         = {2026},
  publisher    = {Zenodo},
  version      = {1.0},
  doi          = {10.5281/zenodo.18868451},
  url          = {https://doi.org/10.5281/zenodo.18868451}
}

@inproceedings{wu2025beneath,
  title={Beneath the Heavens: A Thorough Measurement Study of the Starlink Terrestrial Network},
  author={Wu, Yanbo and Cui, Mingxin and Gou, Gaopeng and Wei, Yuhao and Xiong, Gang and Li, Zhen and Ju, Xinlei and Xia, Wei},
  booktitle={2025 IEEE/ACM 33rd International Symposium on Quality of Service (IWQoS)},
  pages={1--10},
  year={2025},
  organization={IEEE}
}

@inproceedings{garcia2025detailed,
  title={A detailed characterization of starlink one-way delay},
  author={Garcia, Johan and Sundberg, Simon and Brunstrom, Anna},
  booktitle={Proceedings of the 2025 3rd Workshop on LEO Networking and Communication},
  pages={43--49},
  year={2025}
}

@inproceedings{lai2025leocc,
  title={LeoCC: Making Internet Congestion Control Robust to LEO Satellite Dynamics},
  author={Lai, Zeqi and Li, Zonglun and Wu, Qian and Li, Hewu and Li, Jihao and Xie, Xin and Li, Yuanjie and Liu, Jun and Wu, Jianping},
  booktitle={Proceedings of the ACM SIGCOMM 2025 Conference},
  pages={129--146},
  year={2025}
}

@inproceedings{ohs2025phantomlink,
  title={PhantomLink: Emulating virtual end-to-end links on ground and in orbit},
  author={Ohs, Robin and Stock, Gregory F and Fraire, Juan A and Hermanns, Holger and Schmidt, Andreas},
  booktitle={Proceedings of the 2025 Applied Networking Research Workshop},
  pages={39--46},
  year={2025}
}

@techreport{rfc8289,
  author = {Nichols, Kathleen and Jacobson, Van},
  title = {{Controlled Delay Active Queue Management}},
  howpublished = {Internet Requests for Comments},
  type = {RFC},
  number = {8289},
  year = {2018},
  month = {January},
  institution = {IETF},
  doi = {10.17487/RFC8289}
}

@inproceedings{hammer2024starlink,
  title={Starlink performance through the edge router lens},
  author={Hammer, Sarah-Michelle and Addanki, Vamsi and Franke, Max and Schmid, Stefan},
  booktitle={Proceedings of the 2nd International Workshop on LEO Networking and Communication},
  pages={67--72},
  year={2024}
}

@misc{irtt,
  author = {Heist, Pete},
  title = {{irtt: Isochronous Round-Trip Tester}},
  howpublished = {\url{https://github.com/heistp/irtt}},
  publisher = {GitHub},
  note = {Accessed November 17, 2025}
}

@inproceedings{garcia2025tcp,
  title={TCP Congestion Control Performance over Starlink},
  author={Garcia, Johan and Sundberg, Simon and Brunstrom, Anna},
  booktitle={Proceedings of the 2025 Applied Networking Research Workshop},
  pages={70--77},
  year={2025}
}

@misc{StarlinkStories,
  author = {{Starlink}},
  title = {{Starlink Stories}},
  year = {2025},
  howpublished = {\url{https://stories.starlink.com/}},
  note = {Accessed November 17, 2025}
}

@article{garcia2025modeling,
  title={Modeling and predicting starlink throughput with fine-grained burst characterization},
  author={Garcia, Johan and Beckerle, Matthias and Sundberg, Simon and Brunstrom, Anna},
  journal={Computer Communications},
  pages={108090},
  year={2025},
  publisher={Elsevier}
}

@inproceedings{garcia2024finegrained,
  title={Fine-grained starlink throughput variation examined with state-transition modeling},
  author={Garcia, Johan and Sundberg, Simon and Brunstrom, Anna},
  booktitle={2024 19th Wireless On-Demand Network Systems and Services Conference (WONS)},
  pages={69--76},
  year={2024},
  doi={10.23919/WONS60642.2024.10449629},
  organization={IEEE}
}

@inproceedings{Hu23LEOvsCellular,
author = {Hu, Bin and Zhang, Xumiao and Zhang, Qixin and Varyani, Nitin and Mao, Z. Morley and Qian, Feng and Zhang, Zhi-Li},
title = {LEO Satellite vs. Cellular Networks: Exploring the Potential for Synergistic Integration},
year = {2023},
booktitle = {Companion of the 19th International Conference on Emerging Networking EXperiments and Technologies},
pages = {45–51},
series = {CoNEXT 2023}
}

@inproceedings{cao2023satcp,
  title={SaTCP: Link-Layer Informed TCP Adaptation for Highly Dynamic LEO Satellite Networks},
  author={Cao, Xuyang and Zhang, Xinyu},
  booktitle={IEEE INFOCOM 2023-IEEE Conference on Computer Communications},
  pages={1--10},
  year={2023},
  organization={IEEE}
}

@inproceedings{beckman2024starlink,
  title={Starlink and Cellular Connectivity under Mobility: Drive Testing Across the Arctic Circle},
  author={Beckman, Claes and Garcia, Johan and Mikkelsen, Herman and Persson, Patrik},
  booktitle={2024 Wireless Telecommunications Symposium (WTS)},
  pages={1--9},
  year={2024},
  organization={IEEE}
}

@inproceedings{pan2024measuring,
  title={Measuring the Satellite Links of a LEO Network},
  author={Pan, Jianping and Zhao, Jinwei and Cai, Lin},
  booktitle={IEEE International Conference on Communications},
  year={2024}
}

@article{laniewski2024starlink,
  title={Starlink on the Road: A First Look at Mobile Starlink Performance in Central Europe},
  author={Laniewski, Dominic and Lanfer, Eric and Beginn, Simon and Dunker, Jan and D{\"u}ckers, Michael and Aschenbruck, Nils},
  journal={arXiv preprint arXiv:2403.13497},
  year={2024}
}

@article{laniewski2024wetlinks,
  title={WetLinks: a Large-Scale Longitudinal Starlink Dataset with Contiguous Weather Data},
  author={Laniewski, Dominic and Lanfer, Eric and Meijerink, Bernd and van Rijswijk-Deij, Roland and Aschenbruck, Nils},
  journal={arXiv preprint arXiv:2402.16448},
  year={2024}
}

@inproceedings{deutschmann2023internet,
  title={Internet via Satellite: GEO vs. LEO, OpenVPN vs. Wireguard, and CUBIC vs. BBR},
  author={Deutschmann, J{\"o}rg and Jahandar, Saeid and Hielscher, Kai-Steffen and German, Reinhard},
  booktitle={Proceedings of the 1st ACM MobiCom Workshop on Satellite Networking and Computing},
  pages={19--24},
  year={2023}
}

@inproceedings{kassem2022browser,
  title={A browser-side view of {Starlink} connectivity},
  author={Kassem, Mohamed M and Raman, Aravindh and Perino, Diego and Sastry, Nishanth},
  booktitle={22nd ACM Internet Measurement Conference},
  pages={151--158},
  year={2022},
  series ={IMC '22}
}

@inproceedings{ma2023network,
  title={Network Characteristics of {LEO} Satellite Constellations: A {Starlink}-Based Measurement from End Users},
  author={Ma, Sami and Chou, Yi Ching and Zhao, Haoyuan and Chen, Long and Ma, Xiaoqiang and Liu, Jiangchuan},
  booktitle={IEEE INFOCOM Conference on Computer Communications},
  pages={1--10},
  year={2023},
}

@article{humphreys2023signal,
  title={Signal structure of the {Starlink} {Ku}-band downlink},
  author={Humphreys, Todd E and Iannucci, Peter A and Komodromos, Zacharias M and Graff, Andrew M},
  journal={IEEE Transactions on Aerospace and Electronic Systems},
  year={2023},
  publisher={IEEE}
}

@online{starlink2024latency,
  author = {{SpaceX}},
  title = {Improving {S}tarlink's latency},
  url = {https://api.starlink.com/public-files/StarlinkLatency.pdf},
  year = {2024},
  lastaccessed = {June 25, 2024},
}

@inproceedings{garcia2023multi,
  title={Multi-Timescale Evaluation of {Starlink} Throughput},
  author={Garcia, Johan and Sundberg, Simon and Caso, Giuseppe and Brunstrom, Anna},
  booktitle={Proceedings of the 1st ACM Workshop on LEO Networking and Communication},
  pages={31--36},
  year={2023}
}

@inproceedings{garcia2023physRateInference,
  title={Inferring {Starlink} Physical Layer Transmission Rates Through Receiver Packet Timestamps},
  author={Garcia, Johan and Sundberg, Simon and Brunstrom, Anna},
  booktitle={Proceedings of the 25th IEEE Wireless Communications and Networking Conference (WCNC)},
  pages={},
  year={2024}
}

@inproceedings{mohan2024multifaceted,
    author = {Mohan, Nitinder and Ferguson, Andrew E. and Cech, Hendrik and Bose, Rohan and Renatin, Prakita Rayyan and Marina, Mahesh K. and Ott, J\"{o}rg},
    title = {A Multifaceted Look at {Starlink} Performance},
    year = {2024},
    doi = {10.1145/3589334.3645328},
    booktitle = {Proceedings of the ACM on Web Conference 2024},
    pages = {2723–2734}
}

@inproceedings{pan2023measuring,
  title={Measuring a low-earth-orbit satellite network},
  author={Pan, Jianping and Zhao, Jinwei and Cai, Lin},
  booktitle={IEEE 34th Annual International Symposium on Personal, Indoor and Mobile Radio Communications (PIMRC)},
  pages={1--6},
  year={2023},
}

@article{yang2024mobility,
  title={Mobility-aware congestion control for multipath QUIC in integrated terrestrial satellite networks},
  author={Yang, Wenjun and Cai, Lin and Shu, Shengjie and Pan, Jianping},
  journal={IEEE Transactions on Mobile Computing},
  year={2024},
  publisher={IEEE}
}

@inproceedings{fang2024robust,
  title={Robust Live Streaming over LEO Satellite Constellations: Measurement, Analysis, and Handover-Aware Adaptation},
  author={Fang, Hao and Zhao, Haoyuan and Shi, Jianxin and Zhang, Miao and Wu, Guanzhen and Chou, Yi Ching and Wang, Feng and Liu, Jiangchuan},
  booktitle={Proceedings of the 32nd ACM International Conference on Multimedia},
  pages={5958--5966},
  year={2024}
}

@inproceedings{ghafoori2024stars,
  title={Stars and towers on the wheels: Global perspective on satellite networks vs. terrestrial 5G},
  author={Ghafoori, Amirreza and Famili, Alireza and Stavrou, Angelos},
  booktitle={GLOBECOM 2024-2024 IEEE Global Communications Conference},
  pages={301--306},
  year={2024},
  organization={IEEE}
}

@inproceedings{zhang2024starstream,
  title={StarStream: Live Video Analytics over Space Networking},
  author={Zhang, Miao and Li, Jiaxing and Zhao, Haoyuan and Shen, Linfeng and Liu, Jiangchuan},
  booktitle={Proceedings of the 32nd ACM International Conference on Multimedia},
  pages={7909--7917},
  year={2024}
}

@inproceedings{jiang2023leotp,
  title={Leotp: An information-centric transport layer protocol for LEO satellite networks},
  author={Jiang, Li and Zhang, Yihang and Yin, Jinyu and Zhang, Xinggong and Liu, Bin},
  booktitle={2023 IEEE 43rd International Conference on Distributed Computing Systems (ICDCS)},
  pages={579--590},
  year={2023},
  organization={IEEE}
}

@inproceedings{lai2024mind,
  title={Mind the Misleading Effects of LEO Mobility on End-to-End Congestion Control},
  author={Lai, Zeqi and Li, Zonglun and Wu, Qian and Li, Hewu and Liu, Weisen and Liu, Yijie and Xie, Xin and Li, Yuanjie and Liu, Jun},
  booktitle={Proceedings of the 23rd ACM Workshop on Hot Topics in Networks},
  pages={34--42},
  year={2024}
}

@inproceedings{barbosa2023comparative,
  title={A comparative evaluation of TCP congestion control schemes over low-Earth-orbit (LEO) satellite networks},
  author={Barbosa, George and Theeranantachai, Sirapop and Zhang, Beichuan and Zhang, Lixia},
  booktitle={Proceedings of the 18th Asian Internet Engineering Conference},
  pages={105--112},
  year={2023}
}

@article{izhikevich2024global,
  title={A Global Perspective on the Past, Present, and Future of Video Streaming over Starlink},
  author={Izhikevich, Liz and Enghardt, Reese and Huang, Te-Yuan and Teixeira, Renata},
  journal={Proceedings of the ACM on Measurement and Analysis of Computing Systems},
  volume={8},
  number={3},
  pages={1--22},
  year={2024},
  publisher={ACM New York, NY, USA}
}

@inproceedings{kamel2024starquic,
  title={StarQUIC: Tuning Congestion Control Algorithms for QUIC over LEO Satellite Networks},
  author={Kamel, Victor and Zhao, Jinwei and Li, Daoping and Pan, Jianping},
  booktitle={Proceedings of the 2nd International Workshop on LEO Networking and Communication},
  pages={43--48},
  year={2024}
}

@article{ullah2025starlink,
  title={Starlink in Northern Europe: A New Look at Stationary and In-motion Performance},
  author={Ullah, Muhammad Asad and Heikkinen, Antti and Uitto, Mikko and H{\"o}yhty{\"a}, Marko and Anttonen, Antti and Mikhaylov, Konstantin and Lind, Timo},
  journal={arXiv preprint arXiv:2502.15552},
  year={2025}
}

@inproceedings{lanfer2024weather,
  title={Weather-Based Link Prediction for LEO-Satellite Networks using the WetLinks Dataset},
  author={Lanfer, Eric and Laniewski, Dominic and Otten, Daniel and Aschenbruck, Nils},
  booktitle={2024 IFIP Networking Conference (IFIP Networking)},
  pages={586--588},
  year={2024},
  organization={IEEE}
}

@article{laniewski2025measuring,
  title={Measuring Mobile Starlink Performance: A Comprehensive Look},
  author={Laniewski, Dominic and Lanfer, Eric and Aschenbruck, Nils},
  journal={IEEE Open Journal of the Communications Society},
  year={2025},
  publisher={IEEE}
}

@inproceedings{lopez2022empirical,
  title={An Empirical Analysis of Multi-Connectivity between 5G Terrestrial and LEO Satellite Networks},
  author={L{\'o}pez, Melisa and Damsgaard, Sebastian Bro and Rodr{\'\i}guez, Ignacio and Mogensen, Preben},
  booktitle={2022 IEEE Globecom Workshops (GC Wkshps)},
  pages={1115--1120},
  year={2022},
  organization={IEEE}
}

@inproceedings{lopez2023connecting,
  title={Connecting Rural Areas: An Empirical Assessment of 5G Terrestrial-LEO Satellite Multi-Connectivity},
  author={L{\'o}pez, Melisa and Damsgaard, Sebastian Bro and Rodr{\'\i}guez, Ignacio and Mogensen, Preben},
  booktitle={IEEE 97th Vehicular Technology Conference (VTC2023-Spring)},
  pages={1--5},
  year={2023},
}

@article{khan2023realistic,
  title={Realistic assessment of transport protocols performance over LEO-based communications},
  author={Khan, F{\'a}tima and Hervella, Cristina and Diez, Luis and Fern{\'a}ndez, F{\'a}tima and Marcano, N{\'e}stor J Hern{\'a}ndez and Jacobsen, Rune Hylsberg and Ag{\"u}ero, Ram{\'o}n},
  journal={Computer Networks},
  volume={236},
  pages={110008},
  year={2023},
  publisher={Elsevier}
}

@inproceedings{hervella2022realistic,
  title={Realistic Assessment of Transport Protocols Performance over LEO-based Communications},
  author={Hervella, Cristina and Diez, Luis and Fern{\'a}ndez, F{\'a}tima and Hern{\'a}ndez Marcano, N{\'e}stor J and Hylsberg Jacobsen, Rune and Ag{\"u}ero, Ram{\'o}n},
  booktitle={Proceedings of the 19th ACM International Symposium on Performance Evaluation of Wireless Ad Hoc, Sensor, \& Ubiquitous Networks},
  pages={91--98},
  year={2022}
}

@inproceedings{sander2023ecnquic,
  author = {Sander, Constantin and Kunze, Ike and Bl\"{o}cher, Leo and Kosek, Mike and Wehrle, Klaus},
  title = {ECN with QUIC: Challenges in the Wild},
  year = {2023},
  isbn = {9798400703829},
  doi = {10.1145/3618257.3624821},
  booktitle = {Proceedings of the 2023 ACM on Internet Measurement Conference},
  pages = {540–553},
  numpages = {14},
  location = {Montreal QC, Canada},
series = {IMC '23}
}

@article{lim2022ecntraversal,
  title={A Fresh Look at ECN Traversal in the Wild}, 
  author={Hyoyoung Lim and Seonwoo Kim and Jackson Sippe and Junseon Kim and Greg White and Chul-Ho Lee and Eric Wustrow and Kyunghan Lee and Dirk Grunwald and Sangtae Ha},
  year={2022},
  journal={arXiv preprint arXiv:2208.14523},
}

@inproceedings{detal2013tracebox,
  author = {Detal, Gregory and Hesmans, Benjamin and Bonaventure, Olivier and Vanaubel, Yves and Donnet, Benoit},
  title = {Revealing middlebox interference with tracebox},
  year = {2013},
  isbn = {9781450319539},
  publisher = {Association for Computing Machinery},
  address = {New York, NY, USA},
  url = {https://doi.org/10.1145/2504730.2504757},
  doi = {10.1145/2504730.2504757},
  booktitle = {Proceedings of the 2013 Conference on Internet Measurement Conference},
  pages = {1–8},
  numpages = {8},
  location = {Barcelona, Spain},
  series = {IMC '13}
}

@misc{rfc3168,
  series =    {Request for Comments},
  number =    3168,
  howpublished =  {RFC 3168},
  publisher = {RFC Editor},
  doi =       {10.17487/RFC3168},
  url =       {https://www.rfc-editor.org/info/rfc3168},
  author =    {Sally Floyd and Dr. K. K. Ramakrishnan and David L. Black},
  title =     {{The Addition of Explicit Congestion Notification (ECN) to IP}},
  year =      2001,
  month =     sep,
}

@software{nanoping,
  author       = {Yojiro UO},
  title        = {nanoping},
  year         = 2018,
  url          = {https://github.com/iij/nanoping},
  note = {Accessed November 17, 2025}
}

@inproceedings{kargar2016tada,
  title={Tada: An active measurement tool for automatic detection of AQM},
  author={Kargar Bideh, Minoo and Petlund, Andreas and Griwodz, Carsten and Ahmed, Iffat and Behjati, Razieh and Brunstrom, Anna and Alfredsson, Stefan},
  booktitle={Proceedings of the 9th EAI International Conference on Performance Evaluation Methodologies and Tools},
  pages={55--60},
  year={2016}
}

@inproceedings{kua2020detecting,
  title={Detecting bottleneck use of pie or fq-codel active queue management during dash-like content streaming},
  author={Kua, Jonathan and Branch, Philip and Armitage, Grenville},
  booktitle={2020 IEEE 45th conference on local computer networks (LCN)},
  pages={445--448},
  year={2020},
  organization={IEEE}
}

@article{bachl2020detecting,
  title={Detecting Fair Queuing for Better Congestion Control},
  author={Bachl, Maximilian and Fabini, Joachim and Zseby, Tanja},
  journal={arXiv preprint arXiv:2010.08362},
  year={2020}
}

\appendix
\section{Ethics}
This work does not raise any ethical issues.

\section{ECN behavior}
\label{sec:ecn}
We also attempted to investigate if Starlink supports using Explicit Congestion Notification (ECN) to mark packets instead of dropping them. We enable ECN for both a TCP sender and receiver (\texttt{net.ipv4.tcp\_ecn=1}) before establishing a TCP connection over the Starlink connection, but find that the ECN markings appear to be erased over the network path from Karlstad University. We confirm that the ECN usage is negotiated in the TCP handshake. The ECN-Echo (ECE) and Congestion Window Reduce (CWR) bits in the TCP header are appropriately set and outgoing data packets in both directions have the ECN Capable Transport (ECT(0)) code point set in the IP header in accordance with RFC 3168~\cite{rfc3168}. However, all received packets on either side of the connection are marked not-ECT, so the ECN bits are bleached somewhere along the network path.

To determine if the ECN bleaching is caused by Starlink or some other element along the network path, we use a methodology similar to~\cite{sander2023ecnquic,lim2022ecntraversal,detal2013tracebox}. Traceroute is used to send TCP SYN packets with the ECT(0) code point set and ECE flag set and increasing time to live (TTL) values, using the command shown in Listing~\ref{listing:traceroute-insideout-kau}. The returned TTL expired ICMP messages contain a copy of the expired packet header, and by inspecting the ECN bits in the returned header it is possible to infer between which hops the ECN is bleached. From this we find that the ECN markings appear to be able to traverse the Starlink hop, but are erased inside the Twelve99 network connecting us to the Starlink PoP (between hops 8 and 9 in Listing~\ref{listing:traceroute-insideout-kau}).

We later repeat these ECN experiments from the vantage point at the University of Malaga, where we find that ECN markings are preserved across the full end-to-end connection over Starlink. Examining how Starlink handles ECN-marked traffic can therefore be an interesting direction for future work.

\section{Network path}
\label{sec:traceroute}
Listing~\ref{listing:traceroute-insideout-kau} shows the route from the Starlink user terminal at Karlstad University over the satellite link to the Frankfurt PoP, and then back over the terrestrial network to the second NIC port at the measurement server. Listing~\ref{listing:traceroute-insideout-malaga} shows a corresponding trace from the vantage point at the University of Malaga. The traceroute command uses TCP packets with the ECN bits set to ECT(0) to test for ECN bleaching, as described in Appendix~\ref{sec:ecn}, but that has no impact on how the trace is printed here.

\begin{lstlisting}[
    language=sh, 
    float=*, 
    basicstyle=\small\ttfamily, 
    columns=fullflexible, 
    keepspaces=true, 
    breaklines=true, 
    % --- New Visual Settings ---
    backgroundcolor=\color{terminalbg},
    frame=single,                    % Draws a box around the code
    rulecolor=\color{terminalframe}, % Color of the frame
    framesep=8pt,                    % Padding between the text and the frame
    xleftmargin=8pt,                 % Keeps the frame aligned with document margins
    xrightmargin=8pt,
    label={listing:traceroute-insideout-kau}, 
    caption={Traceroute from Starlink terminal at Karlstad University to a public IP address within the university network.}]
$ traceroute -i starlink-uplink -N 1 -t 2 -T -O ecn 193.10.227.25
traceroute to 193.10.227.25 (193.10.227.25), 30 hops max, 44 byte packets
 1  _gateway (192.168.1.1)  0.777 ms  0.769 ms  0.523 ms
 2  100.64.0.1 (100.64.0.1)  30.888 ms  25.297 ms  26.559 ms
 3  172.16.250.10 (172.16.250.10)  26.549 ms  30.666 ms  21.184 ms
 4  * * *
# Starlink PoP (Frankfurt, Germany)
 5  undefined.hostname.localhost (206.224.65.184)  28.401 ms undefined.hostname.localhost (206.224.65.180)  21.823 ms undefined.hostname.localhost (206.224.65.184)  32.280 ms
# Twelve99 (Tier 1 ISP)
 6  ffm-b11-link.ip.twelve99.net (62.115.37.20)  31.677 ms  20.996 ms  23.831 ms
 7  ffm-bb1-link.ip.twelve99.net (62.115.124.116)  21.148 ms  26.354 ms  29.244 ms
 8  kbn-bb5-link.ip.twelve99.net (62.115.143.33)  37.141 ms kbn-bb6-link.ip.twelve99.net (62.115.114.94)  45.043 ms kbn-bb5-link.ip.twelve99.net (62.115.143.33)  31.697 ms
# ECN is bleached past this point
 9  kbn-b4-link.ip.twelve99.net (62.115.134.81)  39.634 ms kbn-b4-link.ip.twelve99.net (62.115.136.231)  37.014 ms  42.782 ms
# NORDUnet/Sunet (Nordic/Swedish academic network)
10  nordunet-ic-358161.ip.twelve99-cust.net (62.115.11.78)  45.201 ms  44.990 ms  37.202 ms
11  mcen1-r11.sunet.se (109.105.102.105)  39.838 ms  42.363 ms  34.522 ms
12  lund-lnd88-r11.sunet.se (86.104.202.102)  37.177 ms  34.276 ms  37.194 ms
13  halmstad-hsd1-r11.sunet.se (86.104.202.48)  45.300 ms  36.903 ms  45.266 ms
14  goteborg-gbg7-r11.sunet.se (86.104.202.28)  39.827 ms  36.983 ms  50.488 ms
15  goteborg-gbg7-r12.sunet.se (86.104.202.35)  42.537 ms  39.622 ms  55.921 ms
16  trollhattan-trh-r11.sunet.se (86.104.202.37)  42.451 ms  45.040 ms  45.223 ms
17  karlstad-kau-r22.sunet.se (86.104.202.159)  42.480 ms  66.543 ms  50.570 ms
18  sunet-gw2.kau.se (130.242.64.121)  47.916 ms  44.792 ms  50.784 ms
19  monroe1.cs.kau.se (193.10.227.25)  42.606 ms  44.955 ms  50.478 ms
\end{lstlisting}

\begin{lstlisting}[
    language=sh, 
    float=*, 
    basicstyle=\small\ttfamily, 
    columns=fullflexible, 
    keepspaces=true, 
    breaklines=true, 
    % --- New Visual Settings ---
    backgroundcolor=\color{terminalbg},
    frame=single,                    % Draws a box around the code
    rulecolor=\color{terminalframe}, % Color of the frame
    framesep=8pt,                    % Padding between the text and the frame
    xleftmargin=8pt,                 % Keeps the frame aligned with document margins
    xrightmargin=8pt,
    label={listing:traceroute-insideout-malaga}, 
    caption={Traceroute from Starlink terminal at University of Malaga to a public IP address within the university network.}]
$ traceroute -i enx5091e353c5df -N 1 -t 2 -T -O ecn www.uma.es
traceroute to www.uma.es (150.214.40.97), 30 hops max, 44 byte packets
 1  customer.mdrdesp1.isp.starlink.com (169.155.233.1)  16.202 ms  18.809 ms  15.650 ms
 2  172.16.251.80 (172.16.251.80)  15.740 ms  18.932 ms  15.712 ms
 3  * * *
# Starlink PoP (Madrid, Spain)
 4  undefined.hostname.localhost (206.224.69.244)  17.418 ms undefined.hostname.localhost (206.224.69.246)  18.254 ms  17.898 ms
# Twelve99 (Tier 1 ISP) 
 5  mad-b3-link.ip.twelve99.net (213.248.89.12)  15.737 ms  19.017 ms  19.705 ms
 6  * * *
 7  be9697.rcr81.b015537-1.mad05.atlas.cogentco.com (154.54.36.170)  16.945 ms  18.994 ms  15.677 ms
 8  149.14.241.18 (149.14.241.18)  15.698 ms  15.161 ms  15.703 ms
# RedIRIS (Spanish academic network)
 9  telmad-rt1.ethtrunk2.cica.rt2.and.red.rediris.es (130.206.245.126)  31.739 ms  23.522 ms  27.574 ms
10  cica-rt2.ethtrunk4.uv.rt2.val.red.rediris.es (130.206.245.34)  31.716 ms ciemat-rt2.ethtrunk2.uv.rt2.val.red.rediris.es (130.206.245.122)  27.025 ms cica-rt2.ethtrunk4.uv.rt2.val.red.rediris.es (130.206.245.34)  31.441 ms
11  rica-backup-router.red.rediris.es (130.206.211.42)  42.960 ms  34.996 ms  31.726 ms
12  uma-router.red.cica.es (150.214.231.170)  35.821 ms  30.991 ms  31.878 ms
13  palo-tea1001.ruma.uma.es (150.214.47.245)  31.541 ms  35.131 ms  31.709 ms
14  fg-sci-tea1001.uma.es (150.214.47.253)  35.580 ms  34.063 ms  35.561 ms
15  ccuma.sci.uma.es (150.214.40.97)  39.695 ms  33.357 ms  31.654 ms
\end{lstlisting}

\section{Example runs randomly selected among runs with loss $>$ 30\%}
\label{sec:additionalRuns}
\begin{figure*}[th]
    \centering
    \includegraphics[width=0.30\textwidth]{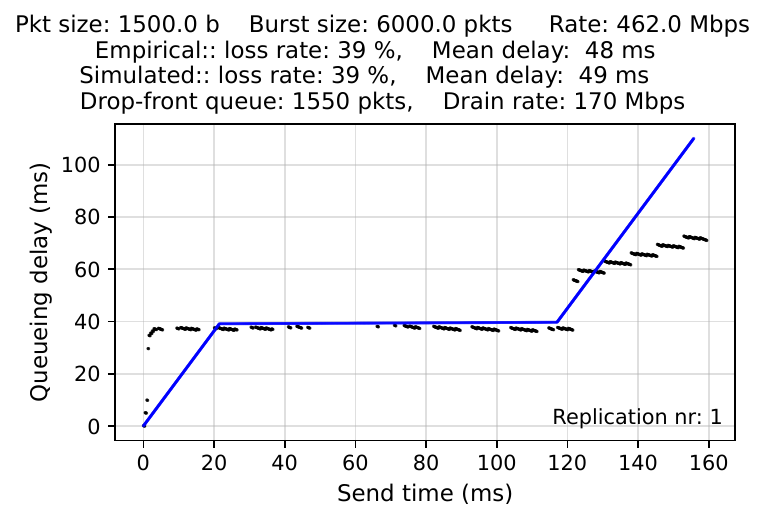}
    \hfill
    \vspace{5mm}
    \includegraphics[width=0.30\textwidth]{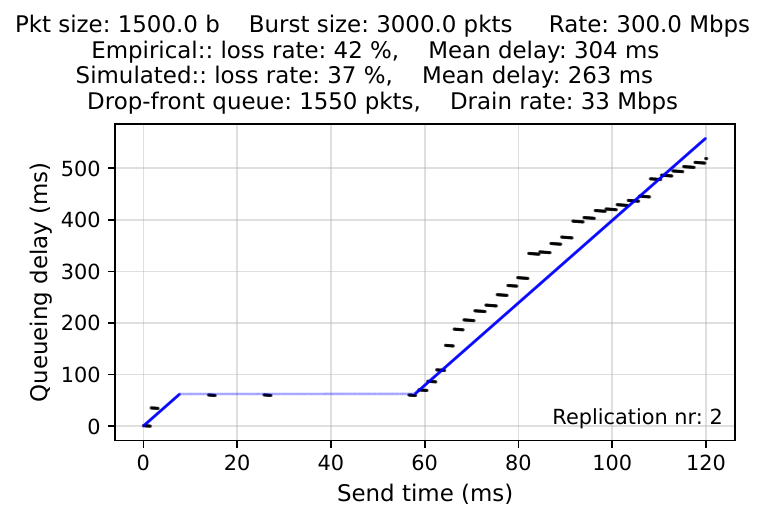}
    \hfill
    \includegraphics[width=0.30\textwidth]{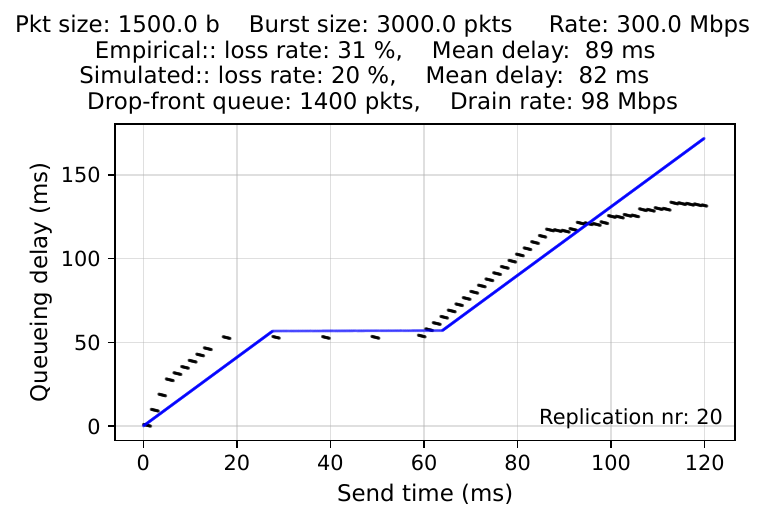}
    \vspace{5mm}
    \includegraphics[width=0.30\textwidth]{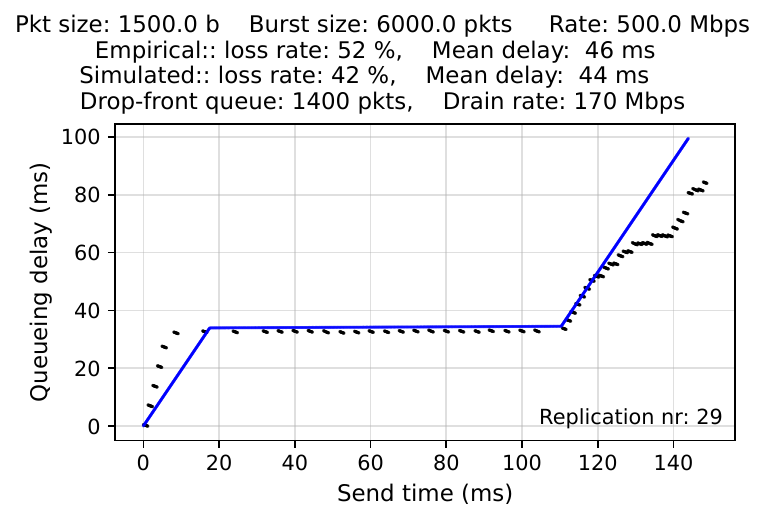}
    \hfill
    \includegraphics[width=0.30\textwidth]{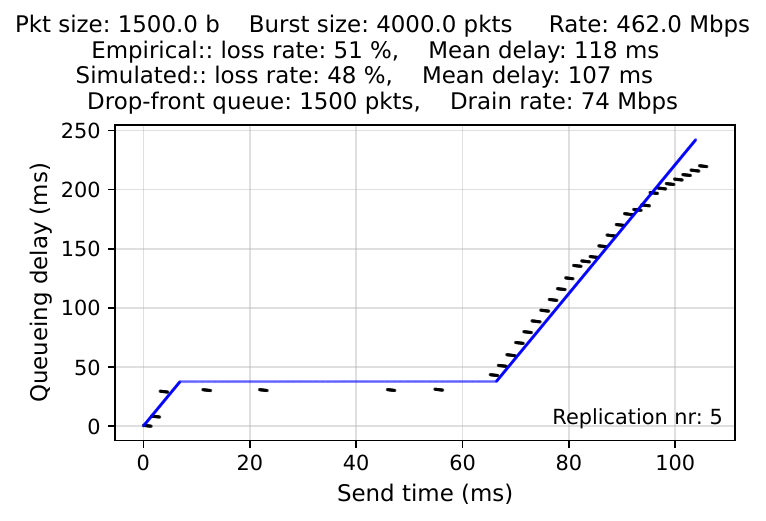}
    \hfill
    \includegraphics[width=0.30\textwidth]{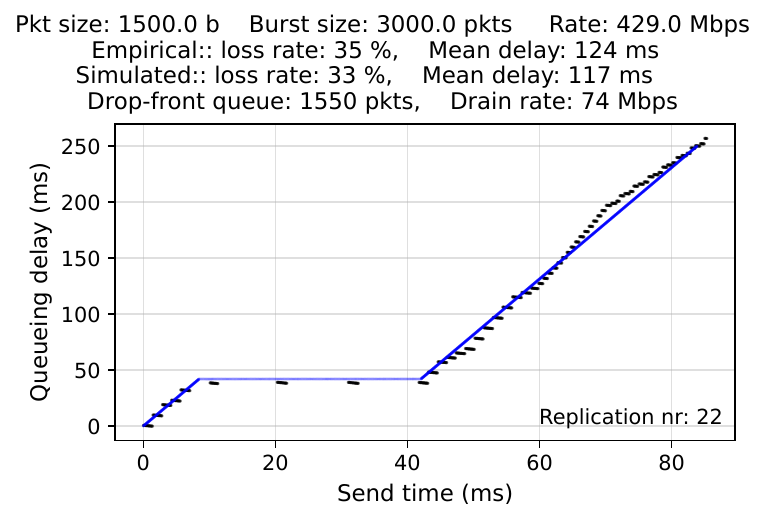}
    \includegraphics[width=0.30\textwidth]{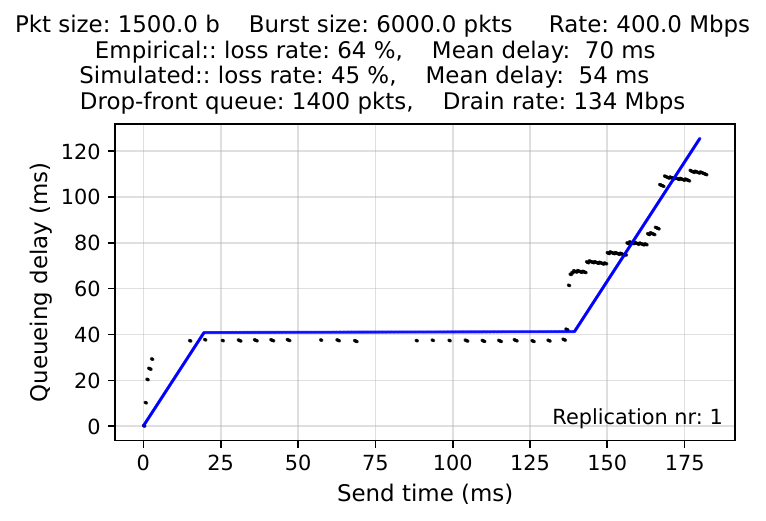}
    \hfill
    \vspace{5mm}
    \includegraphics[width=0.30\textwidth]{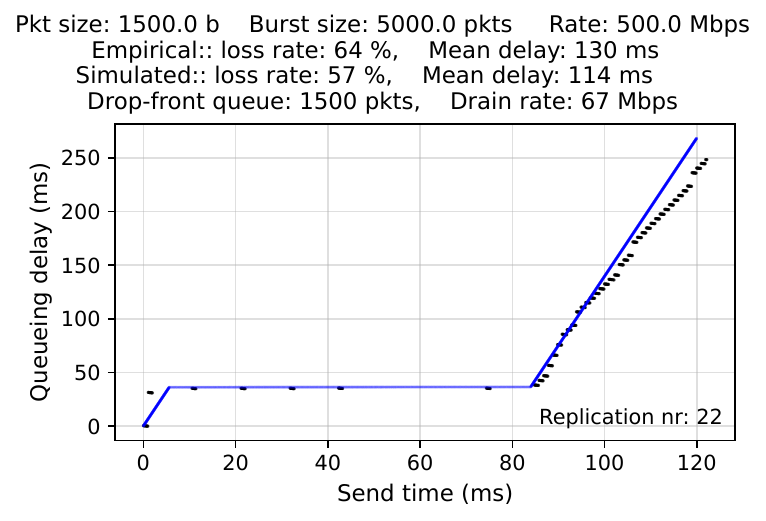}
    \hfill
    \includegraphics[width=0.30\textwidth]{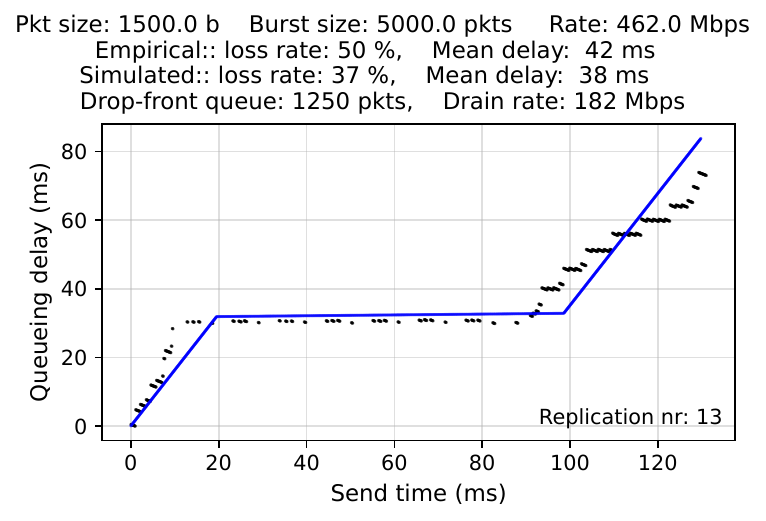}
    \includegraphics[width=0.30\textwidth]{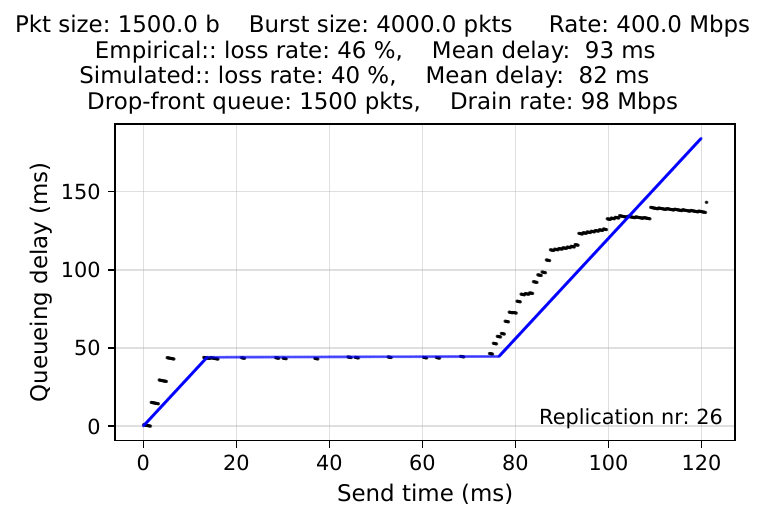}
    \hfill
    \includegraphics[width=0.30\textwidth]{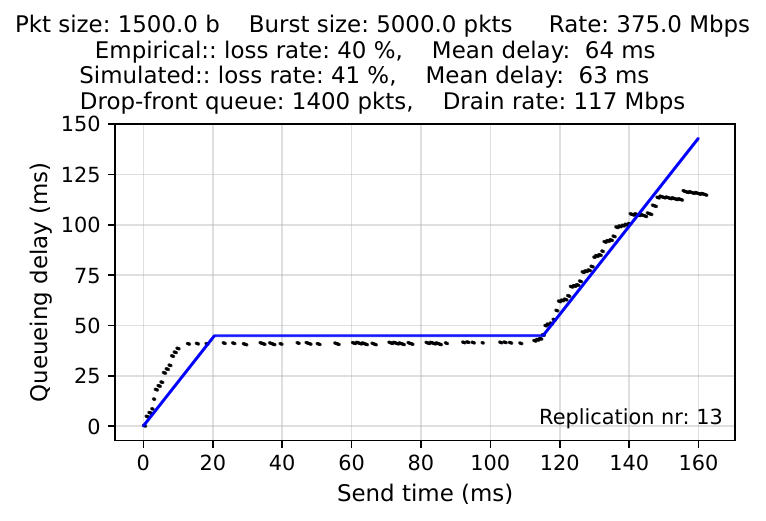}
    \hfill
    \vspace{5mm}
    \includegraphics[width=0.30\textwidth]{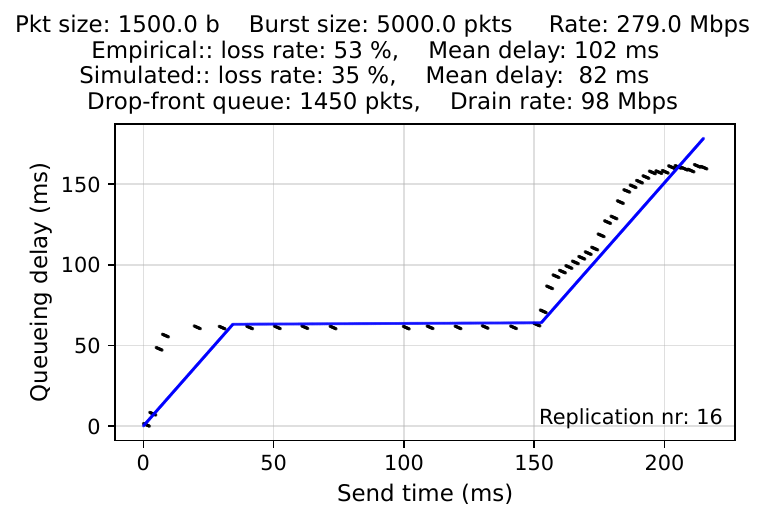}
    \includegraphics[width=0.30\textwidth]{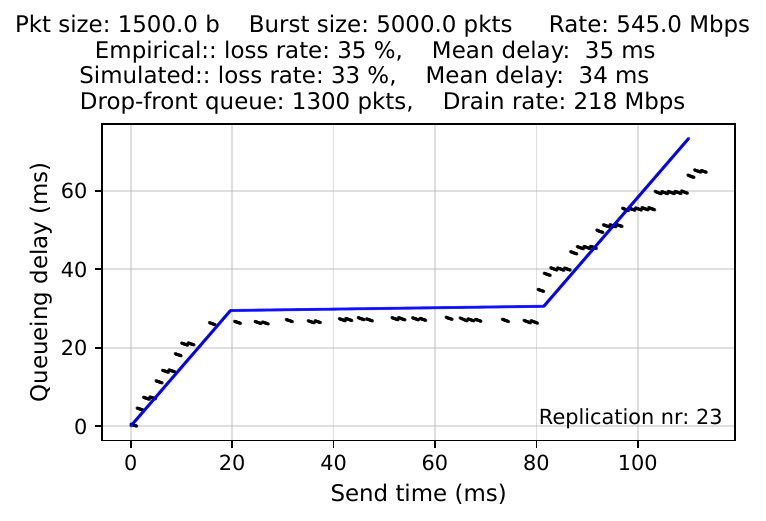}
    \hfill
    \includegraphics[width=0.30\textwidth]{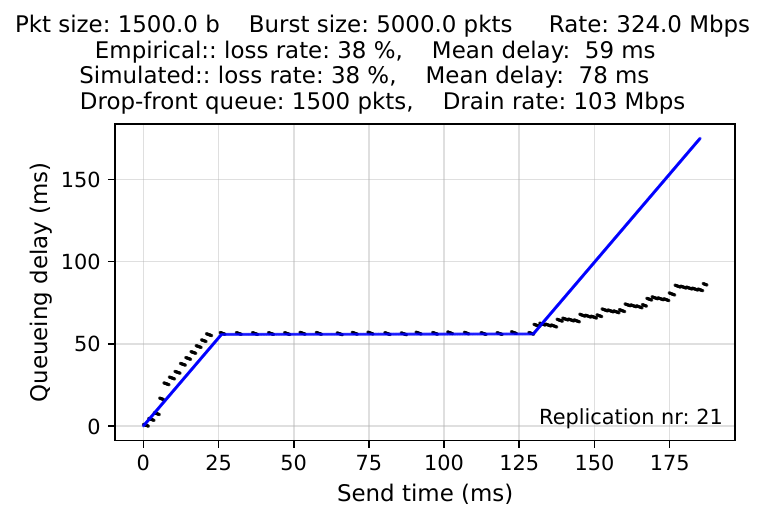}
    \hfill
    \includegraphics[width=0.30\textwidth]{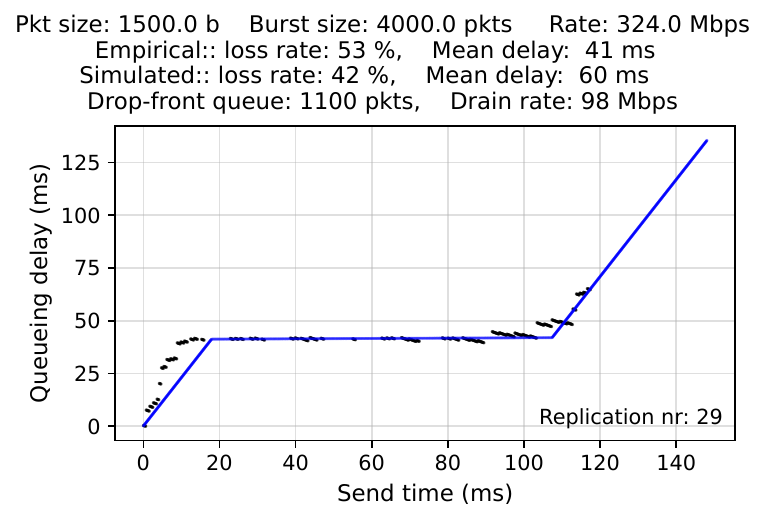}
    \caption{Randomly selected runs showing per-packet queuing-induced delay from measurements (black), and a drop-front queuing simulator configured with queue size and drain rate to fit empirical measurements (blue). }
    \label{fig:examples_simfits}
\end{figure*}

Fifteen randomly selected graphs  are shown in Figure~\ref{fig:examples_simfits} to provide examples of the variation observed during our measurement collection.  On the macroscopic level, it is clear that the empirical behavior is consistent with drop-front buffering.  We found no empirical example consistent with drop-tail. On the microscopic level it is  clear that  the resource allocation patterns vary between runs.  This is also consistent with previous observations of varying resource allocation, i.e. burst patterns~\cite{garcia2025modeling}. 
Several graphs show signs of early rate adaptation  where the empirical slope at the left-hand side of the plateau is steeper than at the right-hand side. For these queuing-delay graphs, such a steeper slope at the beginning indicates a lower initial draining rate. Thus, if the empirical slopes differ on the two sides of the plateau, this suggests that a rate adaptation has taken place.

In the bottom row the middle graph is a clear example of late rate adaptation.  Up to around 130 milliseconds, the simulated and empirically observed behavior fits well. After that, the drain rate increases to approximately 200 Mbps. 
Another example of rate adaptation is present in the right graph of the topmost row. At around 85 milliseconds there is a change in the slope of increase in queuing delay. Since the incoming rate is constant, such a change can only be caused by an increased drain rate. A few other subfigures also have indications of drain rate changes at around the 85 ms time, while others yet have indications of changes at other times. Further work is necessary to better understand the structure of rate adaptations in Starlink.  
The atypical bottom right graph was subject to continuous packet losses towards the end of the run.

\end{document}